\documentclass[final,leqno]{article}
\usepackage[utf8]{inputenc}
\usepackage[english]{babel}
\usepackage{times}
\usepackage{amsmath}
\usepackage{amsfonts}
\usepackage{amssymb}
\usepackage{amsthm}
\usepackage{graphicx}
\usepackage{graphics}
\usepackage{epsfig}
\usepackage{subfig}
\usepackage[section]{placeins}
\usepackage{bm}
\usepackage{fancyhdr}
\usepackage{url}
\usepackage{caption}
\usepackage{booktabs}
\usepackage{cleveref}
\usepackage{float}
\usepackage[letterpaper,margin=1in]{geometry}
\usepackage[colorinlistoftodos]{todonotes}

\newcommand{\comment}[1]{}     % comment{}
\begin{document}
\title{Parameter estimates of the 2016-2017 Zika outbreak in Costa Rica: An Approximate Bayesian Computation (ABC) Approach}
\maketitle

\begin{center}
\author{Fabio Sanchez\footnote{Corresponding author: \url{fabio.sanchez@ucr.ac.cr}\\
Present address: Centro de Investigaci\'on en Matem\'atica Pura y Aplicada (CIMPA), Escuela de Matem\'atica, Universidad de Costa Rica, San Pedro de Montes de Oca, San Jos\'e, Costa Rica, 11501.}, 
Luis Barboza\footnote{Centro de Investigaci\'on en Matem\'atica Pura y Aplicada (CIMPA), Escuela de Matem\'atica, Universidad de Costa Rica, San Pedro de Montes de Oca, San Jos\'e, Costa Rica, 11501. Email: \url{luisalberto.barboza@ucr.ac.cr}} and
Paola V\'asquez\footnote{Escuela de Salud P\'ublica, Universidad de Costa Rica, San Pedro de Montes de Oca, San Jos\'e, Costa Rica, 11501. Email: \url{paola.vasquez@ucr.ac.cr}}}
\end{center}

\begin{abstract}
\noindent In Costa Rica, the first known cases of Zika were reported in 2016. We looked at the 2016-2017 Zika outbreak and explored the transmission dynamics using weekly reported data. A nonlinear differential equation single-outbreak model with sexual transmission, as well as host availability for vector-feeding was used to estimate key parameters, fit the data and compute the {\it basic reproductive number}, $\mathcal{R}_0$, distribution. Furthermore, a sensitivity and elasticity analysis was computed based on the $\mathcal{R}_0$ parameters.
\end{abstract}

%\keywords{
%\textbf{(Zika virus, basic reproductive number, Approximate Bayesian Computation, public health, transmission dynamics, epidemic models, vector-borne diseases)}
%}

%\maketitle
\section{Introduction} \label{sec:intro}
Over the past few decades, arthropod-borne viruses (arboviruses), particularly those transmitted by mosquitoes, have emerged and/or reemerged as significant public health problems worldwide \cite{dash2013}. Different conditions, such as climate change, globalization, human movement, and the lack of adequate vector control strategies, are among the complex factors, that have influenced the growing burden of arboviral infections around the globe \cite{shragai2017}. The present threat that these viruses represent, highlights the importance of a better understating of its transmission dynamics, in order to provide public health care authorities with important evidence that will help in the development of more efficient prevention and control approaches. This research article, focuses on Zika virus (ZIKV), one of the most recent mosquito-borne pathogens that have emerged as a global public health concern \cite{gubler2017}, specifically in the 2016-2017 outbreak that took place in Costa Rica. 

Originally isolated in 1947 from the Zika forest of Uganda \cite{dick1952} and later identified in humans in the 1950's \cite{who2016}, ZIKV remained as a relative obscure arbovirus for nearly 60 years, with fewer than 20 human infections confirmed from countries in Asia and Africa \cite{musso2016,who2015}. It was until 2007, that the disease started to gain global attention, after approximately three quarters of the residents of Yap Island, were infected by the virus \cite{Duffy2009}. This initial outbreak, was followed by a series of epidemics, first in the French Polynesia and other Pacific Islands in 2013-2014, later reaching South America in 2015 \cite{byung2017}. From there, the virus spread rapidly to Central and North America. By January 2018 a total of 223,447 confirmed ZIKV infections, and 3,720 confirmed congenital syndrome associated with ZIKV had been reported to the Pan-American Health Organization \cite{pan2017}. Countries in the Central American Isthmus have reported a total of 7,820 confirmed cases and 190 confirmed congenital syndrome associated with ZIKV \cite{pan2017}. In Costa Rica, during that same period, a total of 10,253 suspected case were reported by the Ministry of Health \cite{mscr}, of which 2,144 were confirmed by laboratory, six newborns and one patient with Guillain-Barr\'e Syndrome have been identified as positive with the ZIKV \cite{inciensa2016,inciensa2017}. However, the actual proportion of human population infected and the number of new cases are extremely difficult to verify \cite{shragai2017}.

In urban environments, ZIKV is mainly transmitted in a human-mosquito-human transmission cycle, with the female \textit{Aedes aegypti} as a primary vector and to a lesser extent \textit{Aedes albopicuts} \cite{Peterson2016,epelboin2016}. However, the recent outbreaks, had led to the identification of multiple confirmed and probable secondary modes of transmission \cite{gregory2017}. It has been documented that the ZIKV, can be passed from a pregnant mother to her child during all trimesters and at time of delivery~\cite{besnard2014,brasil2016,calvet2016}, accidentally as a result of a needle stick injury within a laboratory~\cite{gregory2017} and during sexual contact with an infected partner~\cite{yockey2016}, where the virus has been found in the semen of infected men \cite{musso2015,rowland2016}, even when it is no longer detectable in blood, usually clearing from the semen after about 3 months~\cite{karkhah2018}. The virus, has also been isolated from breast milk \cite{sotelo2017,dupont2016} and in blood transfusions~\cite{bloch2018}. Once a person is infected, the average incubation period is estimated to last between 2 to 7 days~\cite{paho2017}. It has been reported that approximately 80 percent of people infected with the virus are asymptomatic~\cite{Duffy2009}. When symptomatic, the clinical manifestations, are usually mild and self limiting~\cite{habby2018}. The main complications caused by Zika are neurological, where cases of Guillain-Barr\'e syndrome had appeared in infected adults \cite{musso2016}. In neonates infected with Zika, an increased number of nervous system malformations, in particular microcephaly, were reported \cite{mlakar2016}. The diagnosis of the virus, is based on clinical examination, however it is very challenging, because of the overlap of symptoms with other arboviral diseases \cite{barzon2016}, the cross-reaction between antibodies directed to ZIKV and other flaviviruses \cite{moares2018}, and the limited reliable techniques to diagnose infection accurately in those who have symptoms \cite{shragai2017}. To date, an accurate Zika diagnosis is best achieve by real-time RT-PCR \cite{moares2018} and there are no specific treatments available for ZIKV infections, therefore only symptoms such as pain and fever can be treated \cite{paho2017}. 

To discuss potential control strategies of an infectious disease, such as ZIKV, mathematical modelling plays an important role \cite{eisenberg2002}. In general, these models can help decision makers to focus on key aspects of the disease, and are now commonly use in the development of public health policies around the world \cite{Driessche2017,basu2013}. An important threshold quantity that can be estimated with such models, is the {\it basic reproductive number}, $\mathcal{R}_0$ \cite{towers2016}. This threshold can be estimated using a variety of mathematical and statistical techniques, in this article we used an Approximate Bayesian approach (ABC) in order to obtain a parameter distributions and $\mathcal{R}_0$.

Given the data provided by the Ministry of Health of Costa Rica, weekly reported ZIKV cases, and using the statistical methodology described in Section \ref{sec:estimation}, we estimated key model parameters, as well as the $\mathcal{R}_0$ distribution. This, to assess viable public health strategies regarding the key feature of the model, host availability for mosquito feeding.

The article is organized as follows: in Section \ref{sec:methods}, we provide details on the data and statistical methodology that was used to estimate parameters, as well as the description of the model used. In Section \ref{sec:estimation}, we show the parameter estimation using the Approximate Bayesian Computation. In Section \ref{sec:results}, we provide the results and, in Section \ref{sec:conc}, we give our conclusions.
 
\section{Materials and methods} \label{sec:methods}
\subsection{Study area}
Costa Rica, is a country located in the Central America isthmus, bordered by Nicaragua to the North, Panama to the south, the Caribbean Sea to the east and the Pacific Ocean to the south-west. It has a territorial extension of 51,100 square kilometers, politically divided in seven provinces, 82 municipalities and 479 districts. The total population, is approximately 5,003,402 people \cite{inec2018}, of which an estimated 75\% live in urban areas \cite{bancomundial2018}. The country has a natural environment rich in biodiversity, and tropical weather with two well defined seasons, a dry season (December-April) and a rainy season (May-November) \cite{solano2000}. The annual average temperature varies from 27$^{\circ}$C to 10$^{\circ}$C, with an annual rainfall that can go from 1500 liters per squares meter, up to 5000 liters per squares meter, depending on the region \cite{brenes2009}. All tropical conditions that greatly influence \textit{Aedes} mosquito population and behaviour. 

The Ministry of Health, is the main entity responsible for the surveillance of arboviral diseases, and has a long standing experience on its management. In the 1980's the \textit{Ae. aegypti} mosquito re-infested the country, after a brief period of eradication \cite{morice2010}. Since 1993, dengue virus has been endemic in Costa Rica, and three of the four serotypes have circulated the country \cite{mscr}. In 2014, chikungunya virus also emerged, causing outbreaks in different regions of the national territory \cite{mscr}. 

In Costa Rica, Zika was first documented in January 2016, as an imported case from a 25 year old Costa Rican male who contracted the virus after a trip to Colombia \cite{gobcr2016}. Prior to this patient, a 55 year old US tourist that had travelled to Nosara, Guanacaste from 19 to 26 December 2015, was diagnosed with Zika upon his return to Massachusetts \cite{chen2016zika}. Subsequently, in February 2016, the first two autochthonous cases were confirmed by the Ministry of Health, one from a 24 year old pregnant woman, and another from a 32 year old woman, both of them residents of Nicoya, Guanacaste \cite{nacion2016}. Since then, the virus spread rapidly throughout the seven provinces, mainly in areas near the coasts, which are largely infested by the mosquito and have high circulation of the other two arboviruses, dengue and chikungunya. In 2016, the most affected province was Puntarenas, with an incidence rate of 791 cases per 100000, followed by Guanacaste (513 cases per 100000), both of them in the Pacific coast, and Lim\'on (210 cases per 100000) in the Caribbean coast \cite{mscr}. The municipalities that reported the highest incidence rates were: Garabito (3159 cases per 100000 population), Orotina (2900 cases per 100000) and Esparza (2450 cases per 100000), all of them located in the province of Puntarenas. In 2017, transmission continued, however, the number of cases were less than the previous year. Contrary to 2016, the most affected province was Lim\'on (367 cases per 100000), followed by Puntarenas (56 cases per 100000) and Guanacaste (45 cases per 10000) \cite{mscr}. The municipalities with the highest incidence rate were: Siquirres (690 cases per 100000 population), Gu\'acimo (526 cases per 100000 population) and Pococ\'i (395 cases per 100000 population). No notified deaths associated with Zika or concomitant infections by Zika, dengue or chikungunya, have been identified in any of the seven provinces \cite{inciensa2017}.

\begin{figure}[H]
\centering
\subfloat[Zika incidence in Costa Rica, 2016. \label{fig:figure1}]{\includegraphics[width=0.5\linewidth]{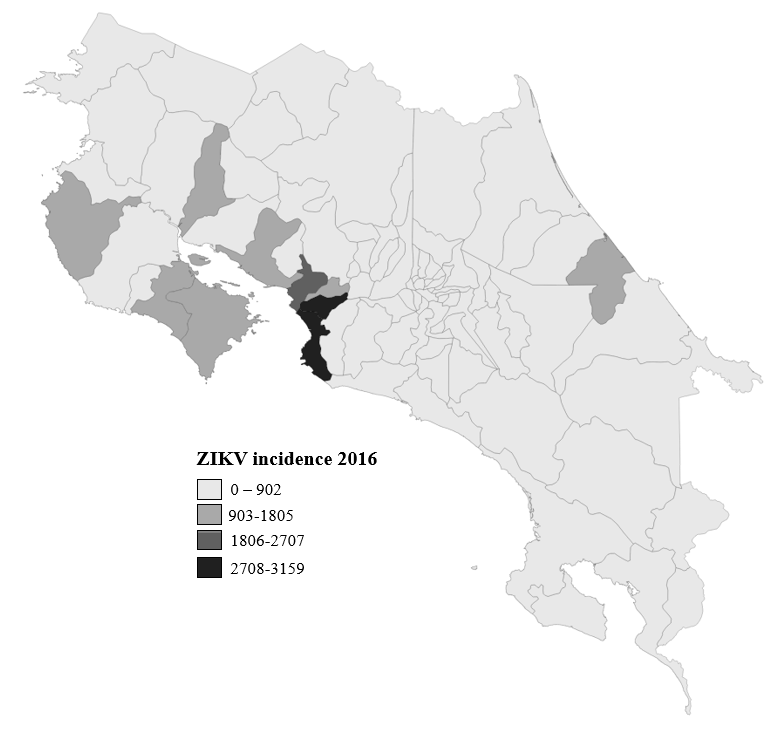}} \hfill
\subfloat[Zika incidence in Costa Rica, 2017. \label{fig:figure2}]{\includegraphics[width=0.5\linewidth]{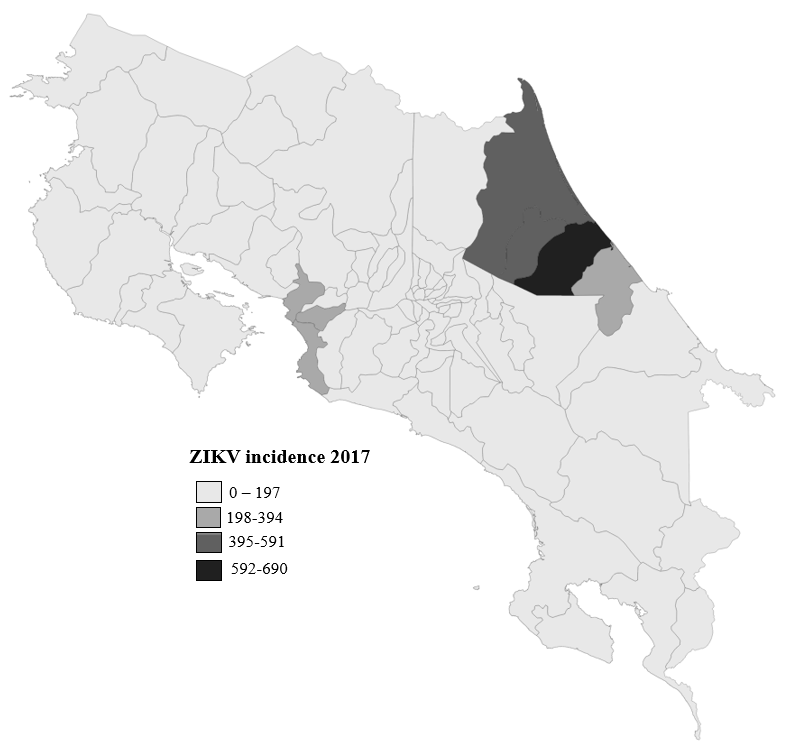}}
\caption{Zika incidence per municipality, Costa Rica, 2016-2017. In 2016, the municipality with higher incidence was Garabito, a municipality located in the Pacific coast of Costa Rica \cite{mscr}. In 2017, the municipality with higher incidence was Gu\'acimo, a municipality located in the Caribbean coast \cite{mscr}.}
\label{fig:map}
\end{figure}
In Figure \ref{fig:map}, we illustrate the concentration of ZIKV cases in 2016 and 2017 in Costa Rica. Note that the vast majority of cases were reported in coastal regions where temperature is ideal for mosquito prevalence.

\subsection{Data}
\begin{figure}[htb!]
\centering
\includegraphics[width=0.7\textwidth]{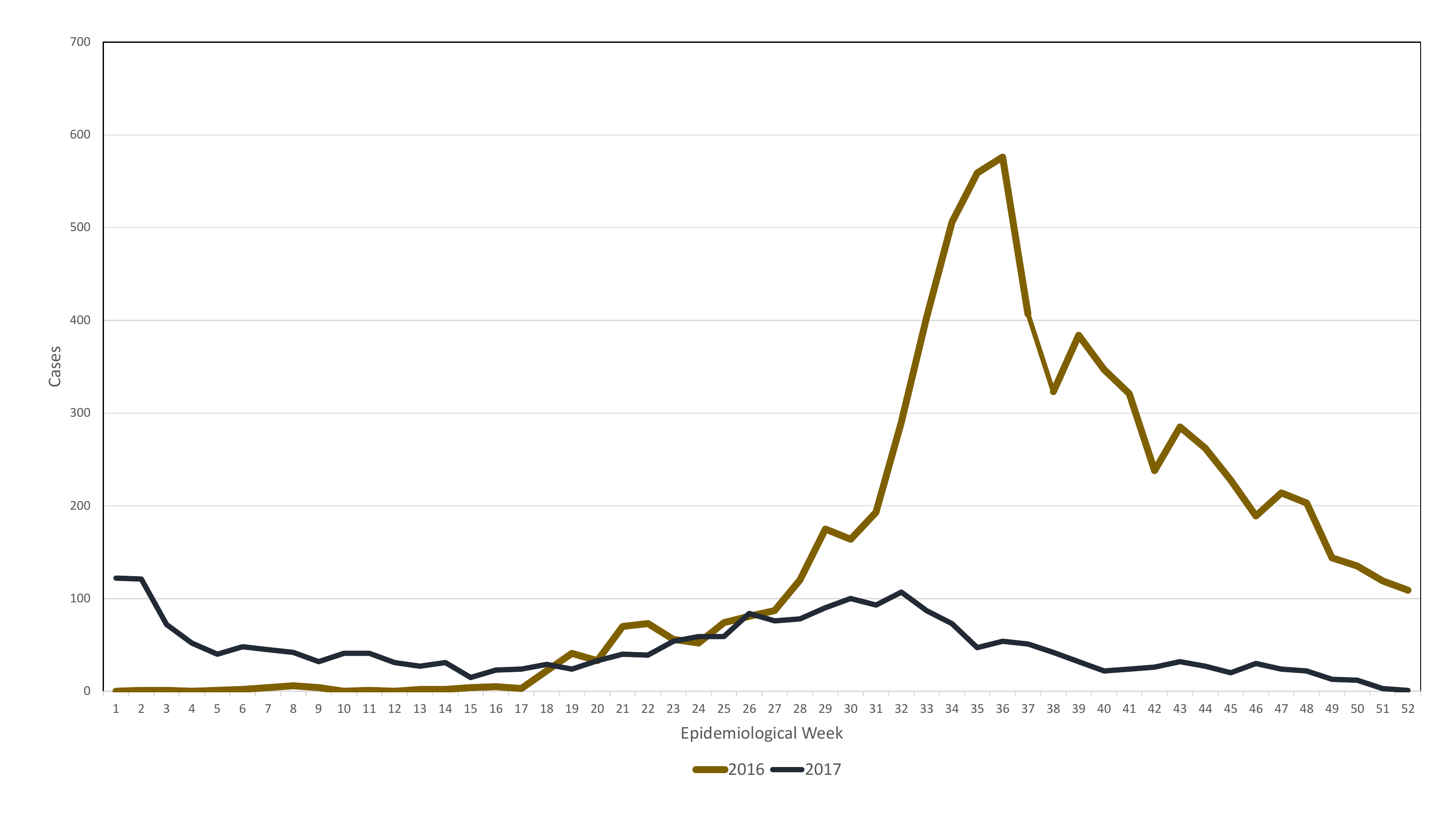}
\caption{Number of reported Zika cases per epidemiological week (EW) in Costa Rica, 2016-2017. In 2016, the Ministry of Health of Costa Rica reported a total of 7,820 cases trough the country, with a peak observed during EW 35 with a total of 605 cases. In 2017, the number of cases were lower, with a total of 2,414 reported cases, the majority of patients were reported during the first EW \cite{mscr}.}
\label{fig:Zcases}
\end{figure}

Weekly reported data of Zika virus cases from 2016-2017 was obtained from the Ministry of Health \cite{mscr}. Between January and December 2016, a total of 7,820 cases were reported from hospitals and clinics around the country. The peak of patients was observed during the months of August and September (epidemiological week 31-41), as shown in Figure \ref{fig:Zcases} \cite{mscr}. That first year, the laboratory responsible for coordinating the virological surveillance of arbovirus at a national level, the INCIENSA National Virology Reference Center, analyzed 6,297 samples, of which 1,794 were positive \cite{inciensa2016}. 

In 2017, transmission continued, however, the number of cases were less than the previous year. A total of 2,414 cases were reported by the Ministry of Health \cite{mscr}, with a confirmation of 350 cases by the INCIENSA National Virology Reference Center \cite{inciensa2017}. 

\subsection{Single outbreak model}
\begin{figure}[htb!]
\centering
\includegraphics[scale=0.5]{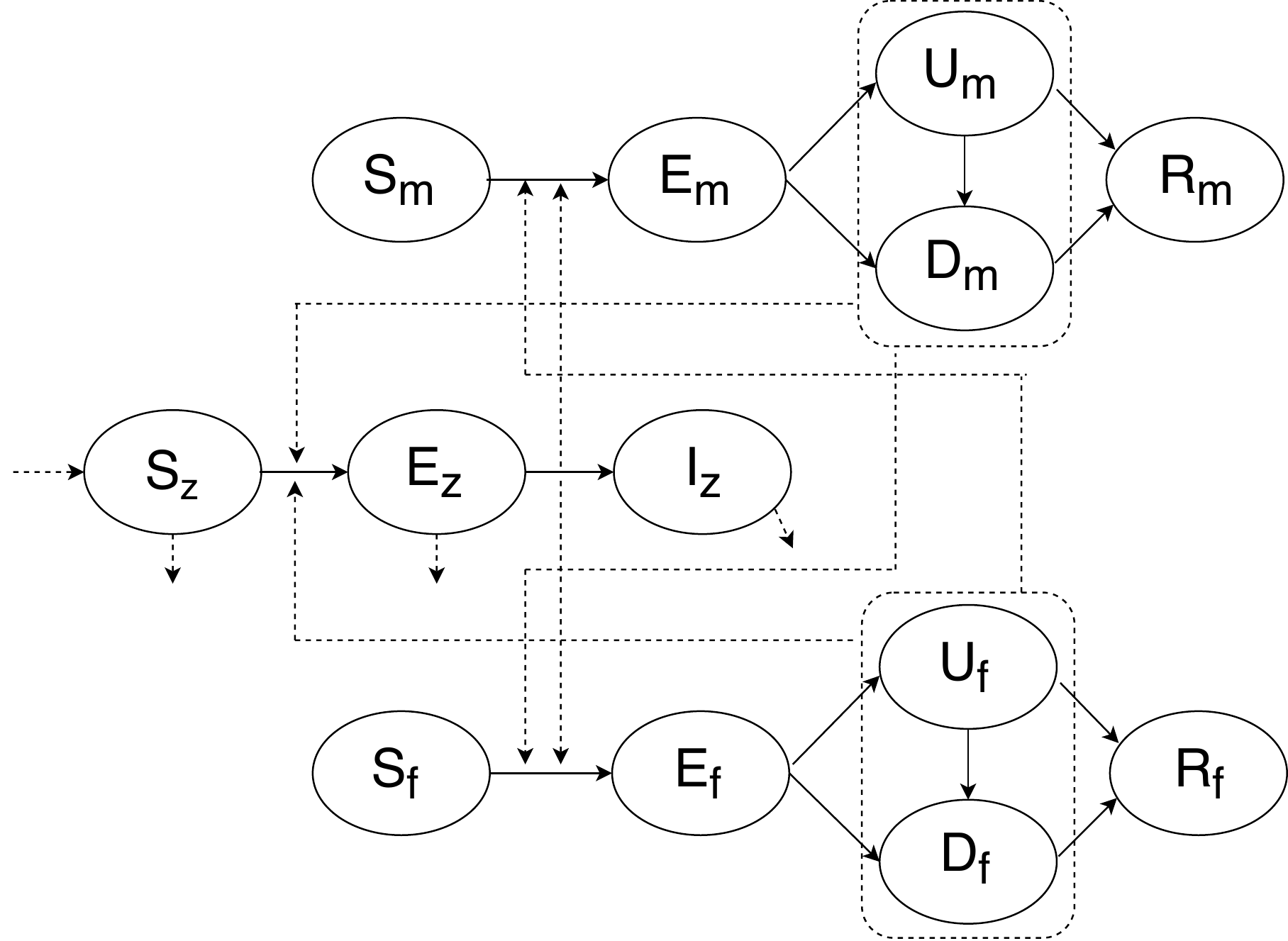}
\caption{Flow diagram of Zika dynamics with sexual transmission.}
\label{fig:model}
\end{figure}

There is an ample history of mathematical models for vector-borne diseases \cite{feng1997,esteva1998,esteva1999,lee2015,brauer2016,sanchez2006,sanchez2012,murillo2014,manore2014,olawoyin2018}, among others. Moreover, other models specifically on estimating parameters for dengue, chikungunya and Zika include: \cite{olawoyin2018,chowell2006,chowell2007,massad2009,towers2015,sanchez2018}. We introduce a nonlinear differential equation single-outbreak epidemic model that describes the Zika dynamics with sexual transmission and host availability for mosquito feeding in Costa Rica. Essentially, individuals who stay home (inside) have a higher chance of being bitten by a mosquito \cite{inec}. In the case of Costa Rica, females conform the smaller proportion of the workforce, hence a larger proportion of females tend to stay at home (inside). In Figure \ref{fig:model}, we illustrate the model transitions.
%\begin{table}[!htb]
  %  \caption{Model variables.}
   % \label{tb:vars}
    %  \centering
     % \scalebox{0.7}{%
      %  \begin{tabular}{|c|c|} \hline
       %     State Variable & Description \\ \hline 
        %    $S_{m,f}$   & Susceptible male and female individuals, respectively\\ 
         %   $E_{m,f}$   & Exposed male and female individuals, respectively (infected but not infectious)\\ 
%            $U_{m,f}$   & Infected (undiagnosed) male and female individuals, respectively\\ 
 %           $D_{m,f}$   & Infected (diagnosed) male and female individuals, respectively\\ 
 %           $R_{m,f}$   & Recovered male and female individuals, respectively\\ 
  %          $S_z$   & Susceptible vectors\\ 
   %         $E_z$   & Exposed vectors (infected but not infectious)\\ 
    %        $I_z$   & Infected vectors\\ \hline
 %      \end{tabular}}
%\end{table}
\begin{table}[!htb]
    \caption{Model parameters and description.}
    \label{tb:pars}
      \centering
      \scalebox{0.7}{%
        \begin{tabular}{|c|c|}\hline
            Parameter & Description \\ \hline 
            $\beta$ & Sexual transmission rate\\ 
            $\sigma_{m,f}$ & Sexual activity probability for males and females, respectively\\ 
            $\alpha_{m,f}$ & Per-capita exposed rate of males and females, respectively\\ 
            $\delta_{m,f}$ & Per-capita diagnosis rate after infection\\ 
            $\gamma_{m,f}$ & Per-capita recovery rate of males and females, respectively\\ 
            $\beta_z$      & Transmission rate (mosquito-human)\\ 
            $\xi_{m,f}$    & Host availability probability of a mosquito feeding on a host\\
            $\mu$          & Per-capita mortality rate of vectors\\
            $\kappa$       & Mosquito biting rate\\
            $p_{vh}$       & Probability of infection from mosquito to human \\ \hline
        \end{tabular}}
\end{table}
The population is divided as follows: $S_{m,f}$-susceptible males and females, respectively, $E_{m,f}$-exposed males and females, respectively (infected but not infectious), $U_{m,f}$-infected (undiagnosed and infectious) males and females, respectively, $D_{m,f}$-infected (diagnosed) males and females, respectively and $R_{m,f}$-recovered (immune) males and females, respectively. For the vector we have: $S_z$ (susceptible mosquitoes), $E_z$ (latent mosquitoes) and $I_z$ (infected mosquitoes). The following is the system of nonlinear differential equations:
\begin{equation}
\label{som}
\begin{array}{rcl}
S_m^\prime &=& -\lambda(t) S_m \frac{I_z}{N_z} - \beta \sigma_m S_m \Big[\frac{U_f+D_f}{N_f}\Big],\\
E_m^\prime &=& \lambda(t) S_m \frac{I_z}{N_z} + \beta \sigma_m S_m \Big[\frac{U_f+D_f}{N_f}\Big]-\alpha_m E_m,\\ 
U_m^\prime &=& \alpha_m E_m - (\gamma_m+\delta_m) U_m,\\
D_m^\prime &=& \delta_m U_m - \gamma_m D_m,\\
R_m^\prime &=& \gamma_m U_m + \gamma_m D_m,\\

S_f^\prime &=& -\lambda(t) S_f \frac{I_z}{N_z} - \beta \sigma_f S_f \Big[\frac{U_m+D_m}{N_m}\Big],\\
E_f^\prime &=& \lambda(t) S_f \frac{I_z}{N_z} + \beta \sigma_f S_f \Big[\frac{U_m+D_m}{N_m}\Big]-\alpha_f E_f,\\ 
U_f^\prime &=& \alpha_f E_f - (\gamma_f+\delta_f) U_f,\\
D_f^\prime &=& \delta_f U_f - \gamma_f D_f,\\
R_f^\prime &=& \gamma_f U_f + \gamma_f D_f,\\

S_z^\prime &=& \mu N_z - \beta_z S_z \Big[\frac{\xi_m(U_m+D_m)+\xi_f(U_f+D_f)}{N}\Big] - \mu S_z,\\
E_z^\prime &=& \beta_z S_z \Big[\frac{\xi_m(U_m+D_m)+\xi_f(U_f+D_f)}{N}\Big] - (\mu+\alpha_z)E_z,\\
I_z^\prime &=& \alpha_z E_z - \mu I_z,\\
\end{array}
\end{equation}

\noindent where $N_m=S_m+E_m+U_m+D_m+R_m$, $N_f=S_f+E_f+U_f+D_f+R_f$, $N=N_m+N_f$ and $N_z=S_z+E_z+I_z$. And the transmission rate is given by $\lambda(t)=\frac{N_z}{N_m+N_f}\kappa p_{hv}$, where $\kappa$ is the mosquito biting rate and $p_{vh}$ is the probability of infection from mosquito to human. Model parameters and description are shown in Table \ref{tb:pars}.

Primarily, we are interested in analyzing the effects of host availability, as well as sexual transmission. Here, the host availability was modeled via the parameter $\xi_{m,f}$ (male and females), where $\xi_{m,f}\in [0,1]$. This parameter serves as a reduction factor in the transmission dynamics from mosquito to host. In the case of sexual transmission, $\sigma_{m,f}$ (males and females), where $\sigma_{m,f}\in[0,1]$, measures the sexual activity in the host population.

Furthermore, using the next generation approach by \cite{heesterbeek1990}, we computed the {\it basic reproductive number}, $\mathcal{R}_0$. However, the analytic expression is highly complex, therefore, we computed its value in {\bf Mathematica} using the parameter estimates.
 
\section{Parameter estimation}\label{sec:estimation}
We used weekly reported cases from~\cite{mscr} and the Approximate Bayesian Computation (ABC) to fit the model, estimate parameters and the {\it basic reproductive number} distribution from the $2016-2017$ Zika outbreak in Costa Rica. 

In Figure \ref{fig:model}, we show the number of Zika confirmed cases and the model solution after estimating parameters. 
 
\subsection{Parameter estimation using ABC} \label{sec:param} 
The parameter estimation method that we use in this article is the Approximate Bayesian Computation (ABC). This method seeks to approximate the posterior distribution of the parameters through algorithms where the evaluation and optimization of the likelihood is not performed. On the other hand, many of these methods are based on sequential algorithms that allow the approximation of the posterior density using sampling schemes such as rejection sampling, MCMC or sequential Monte Carlo sampling (see \cite{Sisson2019} for an exhaustive summary of the subject).

Some of the parameters were kept fixed in order to follow the ecology constraints of the vector, as well as some conditions of the transmission dynamics of the disease that have been studied in previous articles (\cite{manore2014},\cite{sanchez2018}). The total set of parameters (fixed and estimated), as well as the initial conditions of the system are presented in Table \ref{tab:paraminitial}.

The initial conditions of the susceptible populations and the initial values of the three remaining parameters ($\mu$, $\kappa$ and $\delta$) are obtained through the minimization of the sum of squared differences between the observed cases and the total number of cases per week according to the model \eqref{som}. We used a genetic algorithm (\cite{J.Holland}, \cite{GA2013}) and the quasi-Newton L-BFGS-B method (\cite{byrd1995limited}) to obtain the minimum. With the first algorithm we make an initial search in the parameter space and with the second algorithm the search for the first step is improved using as initial value the result of the genetic algorithm.

Once we obtained the initial values of the three parameters, we used a rejection sampling scheme for the ABC (\cite{Pritchard1998}) through the EasyABC package of R (\cite{Jabot2013}). The prior densities for the three parameters were truncated normal to assure that the parameters are strictly positive, and their corresponding means are the initial values obtained above. 

\section{Results}\label{sec:results}
A summary of the main estimation results is shown in Table \ref{tab:paraminitial}. It contains the point estimates and the Bayesian 95\% prediction intervals of the active parameters and the assumed values in case of fixed quantities.  

\begin{figure}[htb]
\centering
\subfloat[$\mathcal{R}_0=1.519$. \label{fig:zikv-a}]{\includegraphics[width=0.3\textwidth]{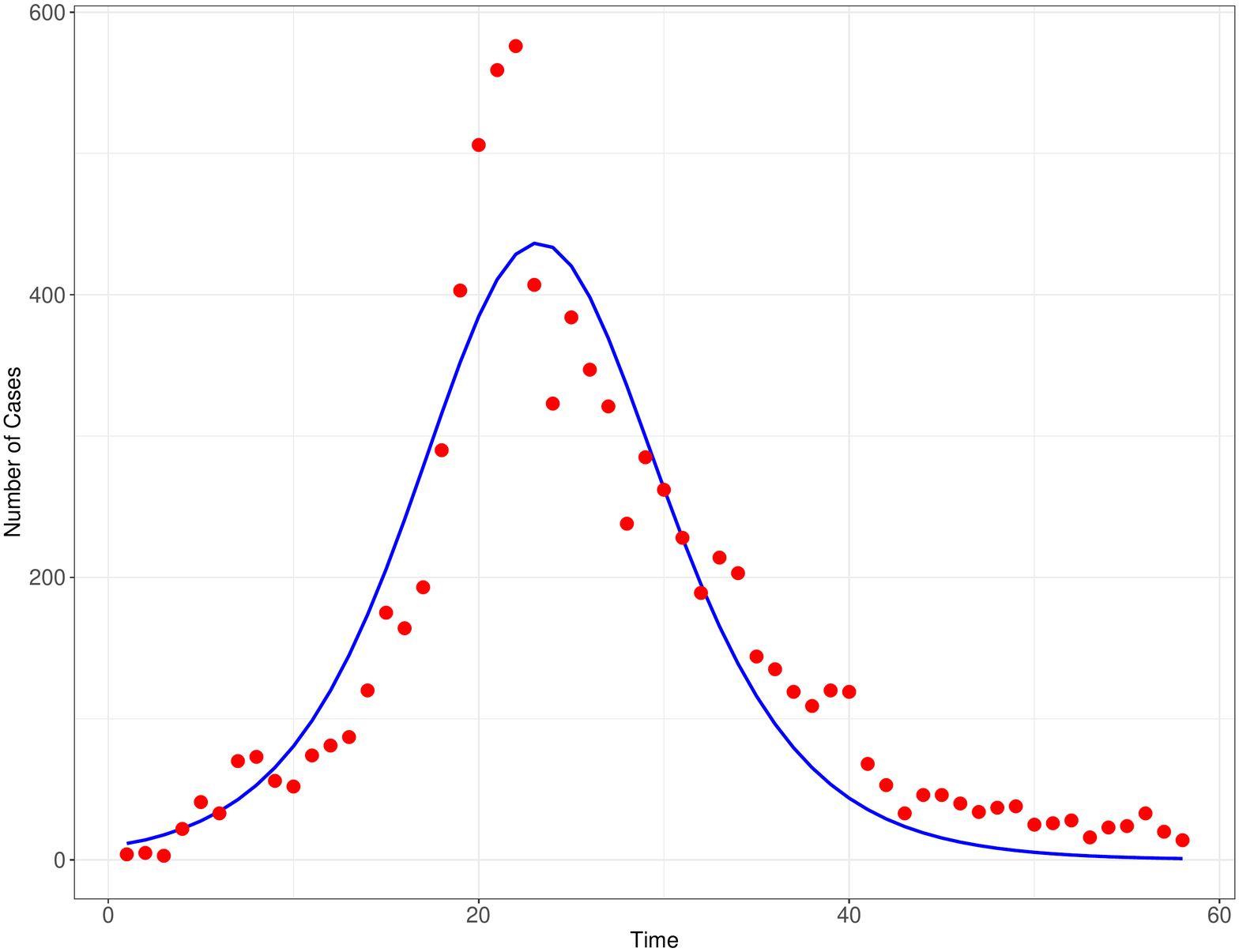}} \hfill
\subfloat[$\mathcal{R}_0=1.434$. \label{fig:zikv-b}]{\includegraphics[width=0.3\textwidth]{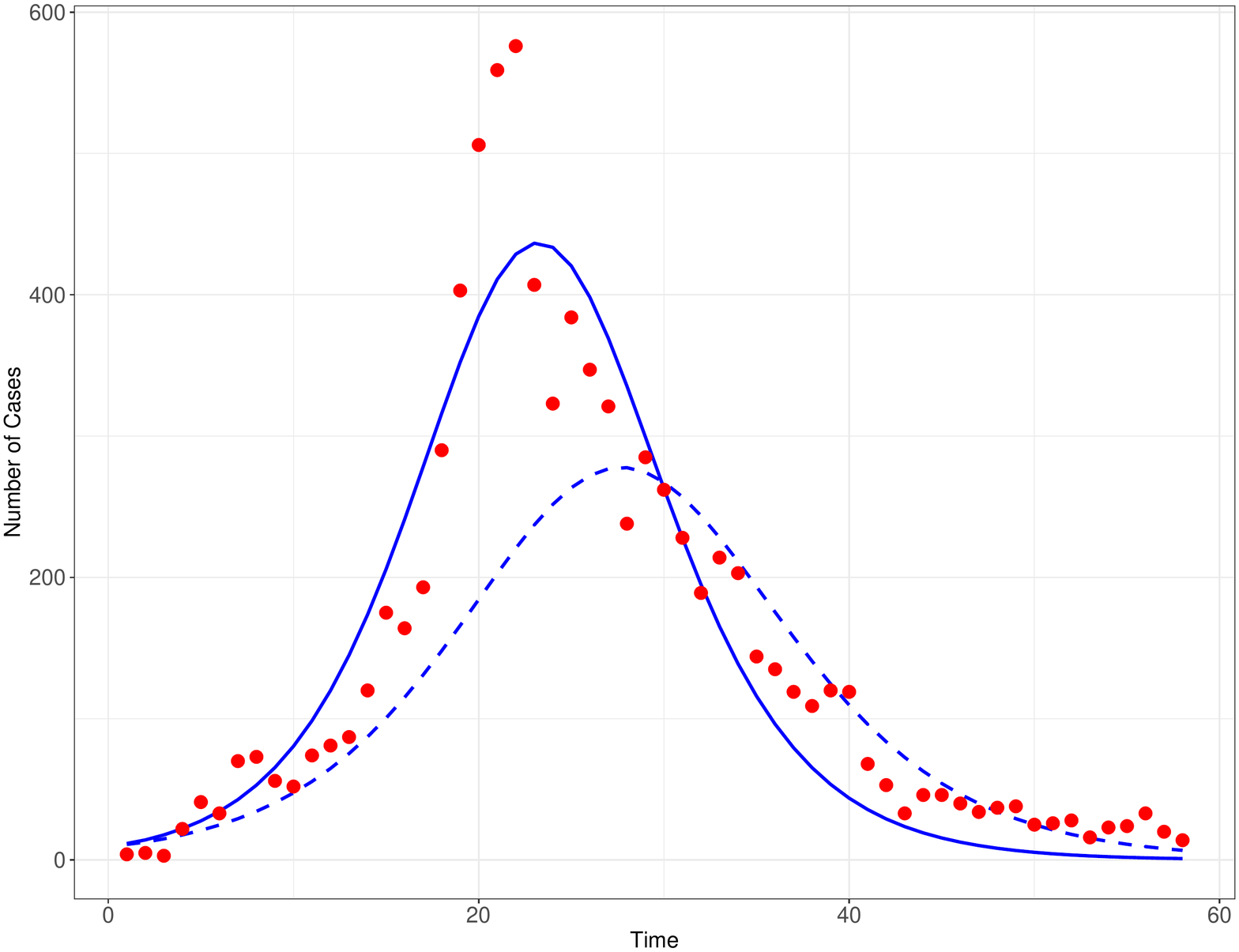}} \hfill
\subfloat[$\mathcal{R}_0=1.368$. \label{fig:zikv-c}]{\includegraphics[width=0.3\textwidth]{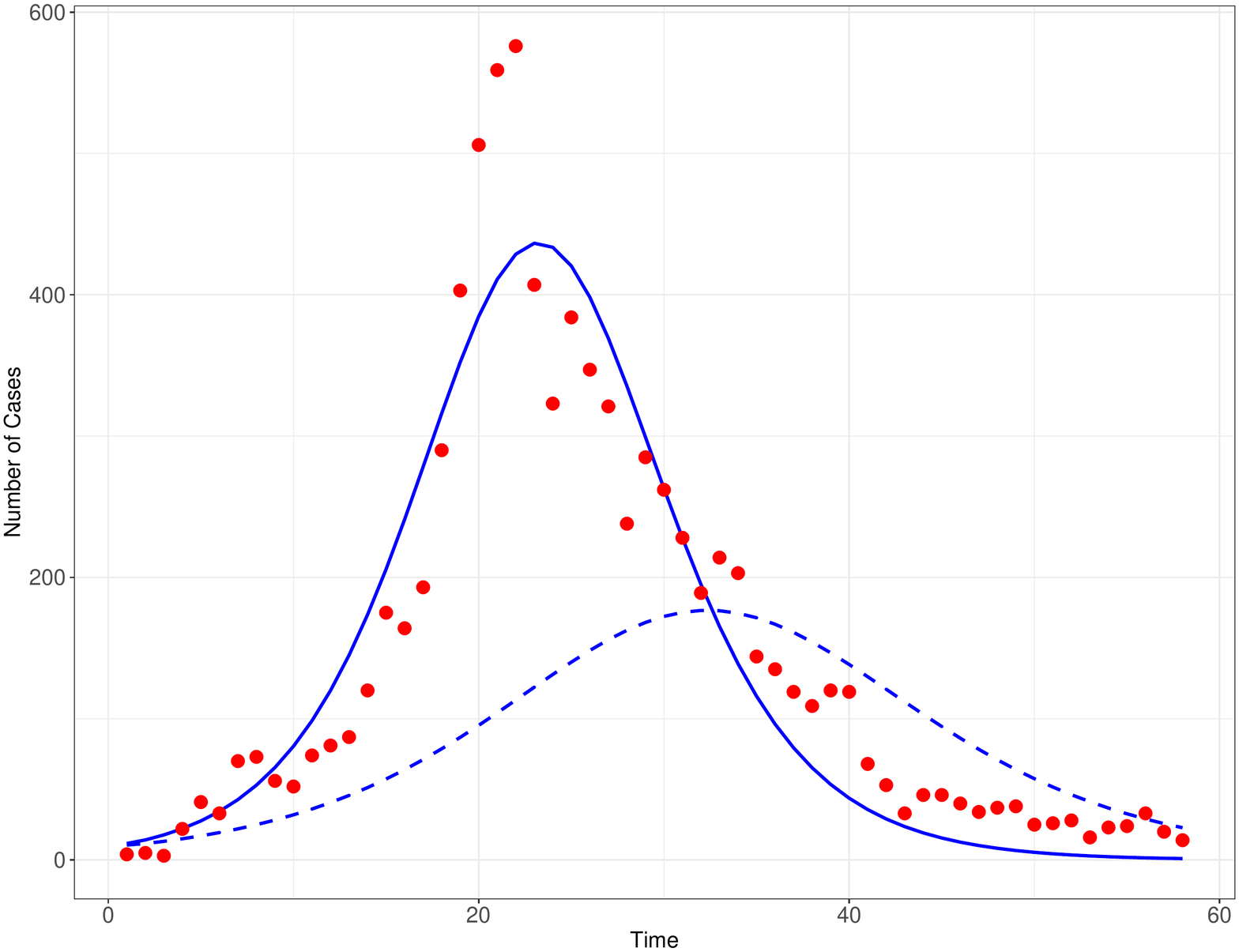}}
\caption{Data (dots) from the 2016-2017 ZIKV outbreak in Costa Rica and model solution (solid line).}
\label{fig:zikaoutbreak}
\end{figure}

In Figure \ref{fig:zikaoutbreak} we show the data and model solutions based on the estimated parameters and initial conditions (see Table \ref{tab:paraminitial}) and different model solutions based on $\xi_m$ and $\xi_f$, as well as the respective $\mathcal{R}_0$ values. In Figure \ref{fig:zikv-a}, the value of $\xi_m=0.5$ and $\xi_f=0.8$. Furthermore, in Figure \ref{fig:zikv-b}, we have $\xi_m=0.5$ and $\xi_f=0.5$, with a reduction in female host availability for mosquito feeding. This leads to a significant reduction in ZIKV cases and an $\mathcal{R}_0=1.434$. In Figure \ref{fig:zikv-c}, a more extreme case where $\xi_m=\xi_f=0.4$, host availability for males and females is reduced, we observe an even lower number of ZIKV cases and an $\mathcal{R}_0=1.368$. 

\begin{figure}[htb!]
  \centering
  \includegraphics[scale=0.45]{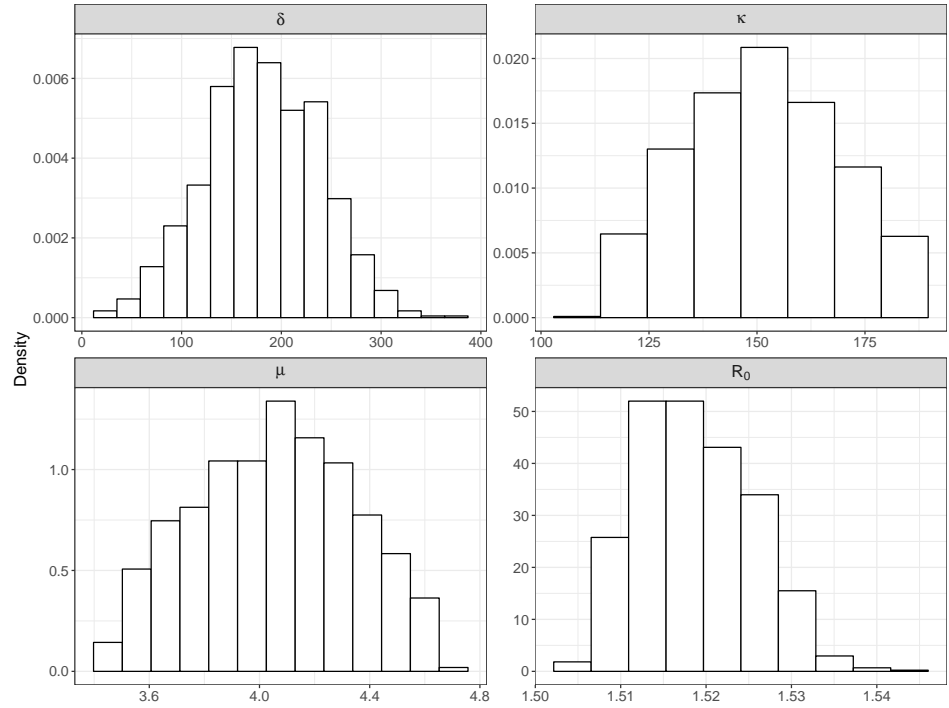}
  \caption{Posterior densities (parameters and $\mathcal R_0$).}
  \label{fig:posteriordens}
\end{figure}

The posterior densities of the three active parameters and the $\mathcal R_0$ are shown in Figure \ref{fig:posteriordens}. We obtained the posterior densities using the ABC approach under a rejection scheme with 100000 samples, from which we selected the 1000 samples that guaranteed the smallest mean square errors between the number of observed and theoretical diagnosed cases. It is important to note that the level of precision of  the parameter estimates, measured though the coefficient of variation of the posterior distribution, attains their minimum values for $\mathcal R_0$ and $\mu$, as can be seen in the last column of Table \ref{tab:paraminitial}. Therefore, the degree of accuracy with which we are estimating $\mathcal R_0$ is largely reliable. 

\begin{table}[H]
  \centering
  \caption{Model parameters, initial conditions and $\mathcal R_0$.}
  \scalebox{0.7}{%
  \begin{tabular}{lcccc}
    \toprule
    \textbf{Parameter} & \textbf{Fixed Value} & \textbf{Point Estimate} & \textbf{Prediction Interval} & \textbf{Coefficient of Variation}\\
    \midrule
    \multicolumn{5}{c}{\textit{Fixed parameters}}\\
    $\alpha_m, \alpha_f$ & 7/5 & N/A & N/A & N/A\\
    $\beta, \beta_z$ & 4.745 & N/A & N/A & N/A \\
    $\gamma_m, \gamma_f$ & 7/6 & N/A & N/A & N/A\\
    $\alpha_z$ & 7/10 & N/A & N/A & N/A\\
    $\sigma_m$ & 0.25 & N/A & N/A & N/A\\
    $\sigma_f$ & 0.2 & N/A & N/A & N/A\\
    $\xi_m$ & 0.5 & N/A & N/A & N/A\\
    $\xi_f$ & 0.8 & N/A & N/A & N/A\\
    $p_{hv}$ & 0.33 & N/A & N/A & N/A\\
    \midrule
    \multicolumn{5}{c}{\textit{Fixed initial conditions}}\\
    $E_m(0),E_f(0)$ & 0 & N/A & N/A & N/A\\
    $U_m(0),U_f(0)$ & 1 & N/A & N/A & N/A\\
    $D_m(0),D_f(0)$ & 0 & N/A & N/A & N/A\\
    $R_m(0),R_f(0)$ & 0 & N/A & N/A & N/A\\
    $E_z(0)$ & 0 & N/A & N/A & N/A\\
    $I_z(0)$ & 1 & N/A & N/A & N/A\\
    \midrule
    \multicolumn{5}{c}{\textit{Active parameters}}\\
    $\kappa$ & N/A & 151.109 & (120.944, 184.707) & 0.116\\
    $\delta_m, \delta_f$ & N/A & 181.307 & (67.337, 288.285) & 0.318\\
    $\mu$ & N/A & 4.054 & (3.539, 4.586) & 0.071\\
    $S_m(0),S_f(0)$ & N/A & 7376 & N/A & N/A \\
    $S_z(0)$ & N/A & 1637 & N/A & N/A\\
    \midrule
    $\mathcal R_0$ & N/A & 1.519 & (1.508, 1.531) & 0.004\\
    \bottomrule
  \end{tabular}}
  \label{tab:paraminitial}
\end{table}

We performed an elasticity and sensitivity analysis on the $\mathcal{R}_0$ parameters (see Table \ref{tb:sens}). Our analysis suggests that host availability, specifically female host availability, is the most sensitive parameter in the model. This suggests viable opportunities for prevention and control strategies. This phenomena is in part due to a large proportion of the female population is out of the workforce. The Costa Rican census \cite{inec} indicates that females old enough to work, that is age 15 or older, approximately 26\% are not in the workforce. This scenario gives the mosquito more feeding opportunities throughout the day assuming they stay home (inside). In contrast, approximately 13\% of the male population is out of the workforce \cite{inec}, in turn, males tend to be more mobile giving the mosquito less opportunities to feed.

\begin{table}[htb!]
\centering
\caption{Sensitivity index for $R_0$ parameters. We used $\frac{\lambda}{R_0}\frac{\partial R_0}{\partial \lambda}$ to determine elasticity ({\it left}) and $\frac{\partial R_0}{\partial \lambda}$ to determine un-normalised sensitivity ({\it right}), where $\lambda$ is the parameter~\cite{arriola2009}. The parameters not shown, $\delta_m$ and $\delta_f$ have indices $<0.001$.}
\label{sensiR0}
\begin{minipage}{0.35\textwidth}
\begin{tabular}{|c|c|}\hline
Parameter & Index     \\ \hline
$\xi_m$    & $28.1998$  \\ \hline
$\xi_f$    & $26.4711$  \\ \hline        
$\beta_z$  & $0.233453$ \\ \hline 
$\kappa$   & $0.233453$ \\ \hline 
$\beta$    & $0.151685$ \\ \hline
$\mu$      & $-0.432531$ \\ \hline 
$\gamma_f$ & $-0.211285$ \\ \hline 
$\gamma_m$ & $-0.172587$ \\ \hline 
\end{tabular} 
\end{minipage}
\begin{minipage}{0.5\textwidth}
\includegraphics[scale=0.2]{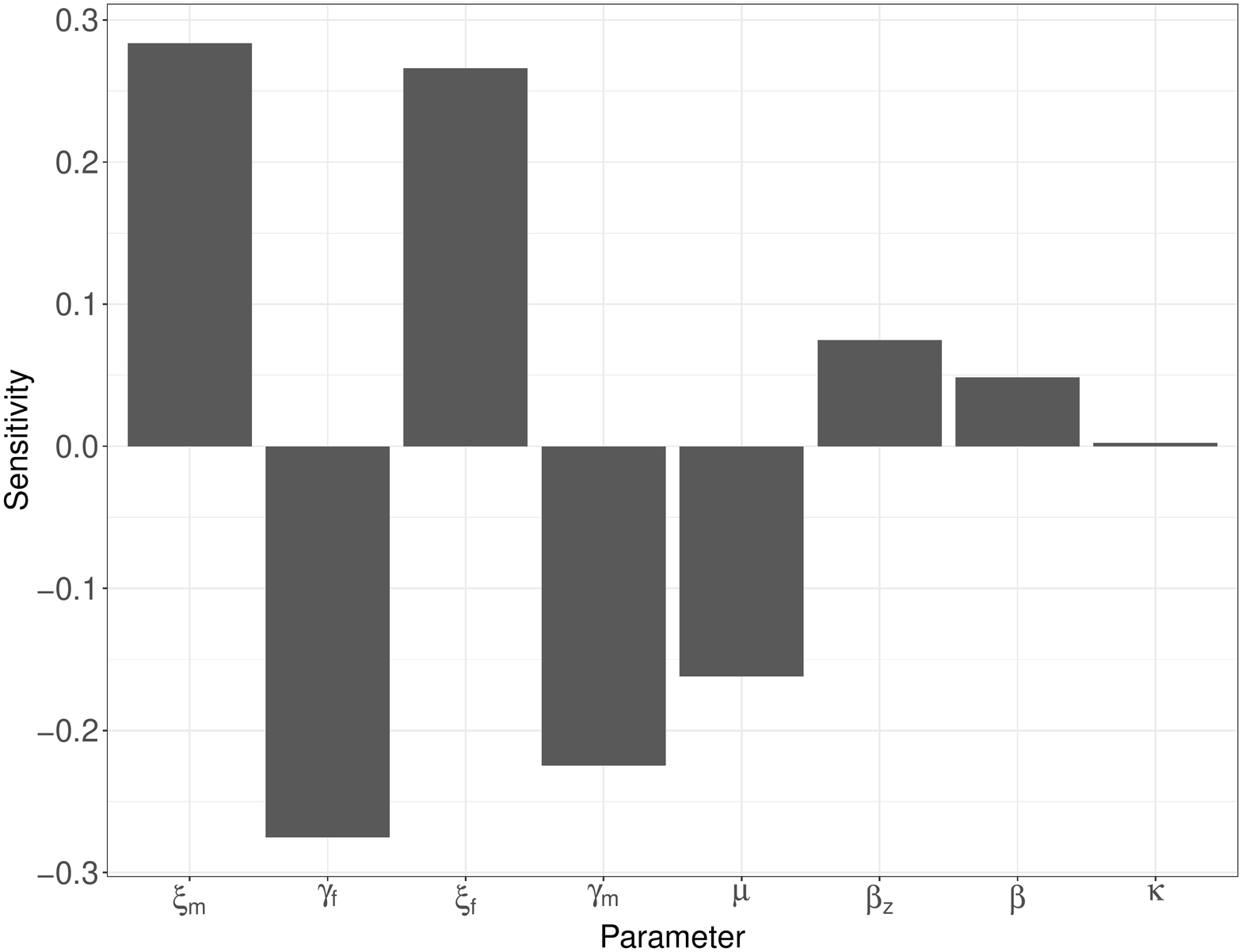}
\end{minipage}
%\quad
%\begin{tabular}{c|c}\hline
%Parameter  & Index  \\ \hline
%$\xi_f$    & $0.266146$  \\ \hline        
%$\xi_m$    & $0.283526$  \\ \hline
%$\beta$    & $0.0485674$ \\ \hline
%$\beta_z$  & $0.0747481$ \\ \hline 
%$\kappa$   & $0.00234718$ \\ \hline 
%$\gamma_f$ & $-0.275143$ \\ \hline 
%$\gamma_m$ & $-0.224749$ \\ \hline 
%$\mu$      & $-0.162096$ \\ \hline 
%\end{tabular}
\label{tb:sens}
\end{table}
%\begin{figure}[htb!]
%\centering
%\subfloat[$\mathcal{R}_0=1.434$.]{\includegraphics[width=0.45\textwidth]{host-b}} \hfill
%\subfloat[$\mathcal{R}_0=1.368$.]{\includegraphics[width=0.45\textwidth]{host-a}}
%\caption{Incidence reduction by host availability in females.}
%\label{fig:avail}
%\end{figure}
We observe that the most sensitive parameters are host availability ($\xi_{m,f}$) followed by the transmission rate from mosquito to human ($\beta_z$) and biting rate ($\kappa$). These parameters increase $\mathcal{R}_0$, therefore creating opportunities to reduce incidence based on strategies in accordance with these parameters may be viable solutions to reduce ZIKV incidence in the future. However, it has been very difficult in the past to significantly reduce the adult mosquito population just by fumigating and other common strategies. In many cases, fumigation is solely done in residential surroundings and not inside homes where female mosquitoes tend to live, close to their food source. Reducing the biting rate could entail strategies surrounding repellents, and in some vulnerable areas this possibility incurs in an economic burden for the at-risk population. Hence, not an optimal or viable solution.  

\section{Conclusions} \label{sec:conc}
After the rapidly spread of ZIKV through different Latin American countries, the introduction of the virus to Costa Rica was expected \cite{minsa2016}. Despite efforts made from public health authorities, the virus has circulated the seven provinces and has caused important consequences in the health and well-being of the Costa Rican population, in particular, due to neurological complications. According to data provided by the Costa Rican Social Security Fund (CCSS), from January 2016-December 2017 there were a total of 21 hospitalizations due to Zika, 13 of which were from women in a reproductive age. During that same period of time, 323 pregnant women and six newborns with microcephaly were confirmed to have the virus \cite{inciensa2018}. However, the actual number of newborns affected by ZIKV could be higher, as evidenced in a report published by The Center for Congenital Diseases in Costa Rica, where they established that since the introduction of ZIKV to the country, the cases of microcephaly almost doubled, and by 2017, the number exceeded almost four times, when compared to the period used as the base line by the Center (2011-2015), affecting all seven provinces in different proportions, being the most affected Guanacaste, Puntarenas and Lim\'on, which could indicate that the observed increase may be associated, in some level, with ZIKV \cite{CREC2017}.   

It is evident that ZIKV possesses an unique challenge to public health authorities around the world. The multiple routes of transmission, the large number of asymptomatic cases, as well as, its link with microcephaly and other neurological disorders, forever changed the public health approach to this virus \cite{carlson2018,honein2018}. In Costa Rica, a country greatly infested by the \textit{Aedes} mosquito, a better understanding of the transmission dynamics,  can provide a guide to introduce more efficient prevention and control interventions, that allows a better use of the human and economic resources available for the control of vector borne diseases. 

Despite not having an analytic expression for the {\it basic reproductive number}, we estimated its distribution using the estimated parameters from Table \ref{tab:paraminitial} by means of a simulation-based Bayesian approach (ABC). This method has the advantage that it does not require strong assumptions on the data which are necessary if we use classical estimation methods, such as maximum likelihood or least squares. Furthermore, the Bayesian nature of ABC allows us to assume that some parameters of the model are random with the additional advantage that the modeling of those parameters includes the understanding of their variation.  

Using data from the Ministry of Health's surveillance system, our study showed that the risk of ZIKV transmission in Costa Rica is greater in the population that spends most of its time inside. This results go in hand with previous studies that evidenced that females adult \textit{Ae. aegypti} disperse relatively short distances \cite{harrington2005}, and highlights the need for more effective community engagement strategies, in order to enable residents of different areas in the communities to make informed health decisions that will influence their overall well-being \cite{juarbe2018}. This becomes more significant, based on the fact that women constitute the majority of the population that stays inside, which gives an extra layer of relevance to the findings. The study also evidenced that for Costa Rica, the sexual transmission route for the virus plays a secondary role in the dispersal of the disease, however, public health officials need to remain vigilant and provide information to the general population about the risk of sexual transmission of ZIKV. 

Although participatory approaches from the communities have been used for many years in the control of mosquito borne diseases, its proper implementation has been difficult to achieve \cite{winch1992}. A bottom-up communicative approach at all levels of the public health system \cite{mustafa2018} could provide a greater interest in general population to implement the mosquito control strategies widely recommended by the public officials around the country. A better understanding of the specific needs of each region, each one with different social, cultural and economic characteristics, makes it fundamental to plan mosquito control strategies according to its specific needs \cite{mena2011}. Not taking into account the great heterogeneity of houses and neighborhoods where the mosquito completes its life cycle, along with the limited resources and personnel trained in the control of the vectors, are at least in part, attributed to the failure of previous prevention and control actions \cite{barrera2000}. Because of the ongoing transmission, and the risk of recurrence of an outbreak, health care authorities around the country need to remain vigilant and establish a comprehensive arboviral disease surveillance. An education and prevention control interventions adjusted to each specific region, with a more active involvement of the communities members is recommended to achieve a more efficient control of the ZIKV and other mosquito-borne diseases. 

\section*{Acknowledgments}
The author is solely responsible for the views and opinions expressed in this research. We would like to thank the Ministry of Health of Costa Rica for providing the data regarding reported Zika cases. 

\section*{Conflict of interest}
All authors declare no conflicts of interest in this paper.

\FloatBarrier

\end{document}